\documentclass[prb,superscriptaddress,twocolumn,floatfix,letterpaper]{revtex4}
\def\officialQ{1}

\usepackage{amssymb, amsmath, mathrsfs}
\usepackage[usenames]{xcolor}
\usepackage{subfigure, wrapfig}
\usepackage{threeparttable}
\usepackage[dvips]{graphicx}

\definecolor{darkgray}{rgb}{0.7,0.7,0.7}
\definecolor{darkred}{rgb}{0.7,0,0}
\definecolor{darkgreen}{rgb}{0,0.7,0}
\definecolor{darkblue}{rgb}{0,0,0.5}
\definecolor{ultralightgray}{rgb}{0.1,0.1,0.1}
\definecolor{lightred}{rgb}{1,0.5,0.5}
\definecolor{lightgreen}{rgb}{0.5,1,0.5}
\definecolor{lightblue}{rgb}{0.5,0.5,1}

\newcommand{\punct}[1]{\textrm{ #1}}
\newcommand{\txt}{\textnormal}

\ifx\officialQ\undefined
	\newcommand{\remark}[1]{{\color{blue} (#1)}}

	\newcommand{\modified}[1]{{\color{magenta} #1}}
	\newcommand{\deleted}[1]{\color{brown} \sout{#1}}
\else
	\newcommand{\remark}[1]{}

	\newcommand{\modified}[1]{#1}
	\newcommand{\deleted}[1]{}
\fi
\ifx\printedBW\undefined
	\newcommand{\colorfig}{}
\else
	\newcommand{\colorfig}{(Color online) }
\fi

\newcommand{\vv}[1]{\mathbf{#1}}
\newcommand{\vvsym}[1]{\boldsymbol{#1}}
\newcommand{\uv}[1]{\hat{#1}}

\newcommand{\tr}{\operatorname{tr}}

\newcommand{\up}{\uparrow}
\newcommand{\dn}{\downarrow}
\newcommand{\dg}{\dagger}
\newcommand{\ket}[1]{| #1 \rangle}
\newcommand{\bra}[1]{\langle #1 |}

\newcommand{\E}{\mathcal{E}}

\newcommand{\tchi}{\tilde{\chi}}
\newcommand{\SdotS}[2]{\vv{S}_{#1} \cdot \vv{S}_{#2}}
\newcommand{\SSS}[3]{\vv{S}_{#1} \cdot (\vv{S}_{#2} \times \vv{S}_{#3})}

\newcommand{\ei}[2][]{\ensuremath #1 e^{#2}_{i}}
\newcommand{\ef}[2][]{\ensuremath #1 \overline{e}^{#2}_{f}}
\newcommand{\efei}[3][]{\ensuremath #1 \overline{e}^{#2}_{f} e^{#3}_{i}}
\newcommand{\cc}[3][]{(#1 c^\dg_{#2} c_{#3})}

\usepackage[dvips]{hyperref}

\ifx\filesasinput\undefined\def\filesasinput{1}\fi
\ifx\usetikzsymbols\undefined\def\usepstsymbols{1}\fi
\usepackage{pstricks}
\ifx\checkpstsymbols\undefined\def\checkpstsymbols{1}\fi
\ifx\checktikzsymbols\undefined\def\checktikzsymbols{1}\fi

\ifx\filesasinput\undefined
\documentclass{article} 
\ifx\usetikzsymbols\undefined\def\usetikzsymbols{1}\fi
\ifx\usepstsymbols\checkpstsymbols\usepackage{pstricks}\fi
\ifx\usetikzsymbols\checktikzsymbols\usepackage{tikz}\fi
\fi

\ifx\usepstsymbols\checkpstsymbols

\newcommand{\wwtov}{\raisebox{-10pt}{
	\psset{unit=0.3 mm}
	\psset{linewidth=1.5,dotsep=1}
	\psset{dotsize=0.7 2.5,dotscale=1 1,fillcolor=black}
	\psset{arrowsize=1 2,arrowlength=1,arrowinset=0.25}
	\begin{pspicture}(14,4)(26,34)
		\psdots[](16,4)(26,19)(16,34)
		\psline[linewidth=1](16,4)(26,19)(16,34)
		\psline[linewidth=1,linestyle=dotted](16,34)(16,4)
		\psline[linestyle=none]{->}(16,4)(21,11)
		\psline[linestyle=none]{->}(26,19)(21,27)
		\psline[linestyle=none]{->}(16,34)(16,20)
	\end{pspicture}
	}
}

\newcommand{\wtovv}{\raisebox{-10pt}{
	\psset{unit=0.3 mm}
	\psset{linewidth=1.5,dotsep=1}
	\psset{dotsize=0.7 2.5,dotscale=1 1,fillcolor=black}
	\psset{arrowsize=1 2,arrowlength=1,arrowinset=0.25}
	\begin{pspicture}(12,4)(24,34)
		\psdots[](24,4)(14,19)(24,34)
		\psline[linewidth=1](24,4)(14,19)(24,34)
		\psline[linewidth=1,linestyle=dotted](24,34)(24,4)
		\psline[linestyle=none]{->}(14,19)(20,10)
		\psline[linestyle=none]{->}(24,34)(19,26)
		\psline[linestyle=none]{->}(24,4)(24,18)
	\end{pspicture}
	}
}

\newcommand{\vtoxx}{\raisebox{-5pt}{
	\psset{unit=0.3 mm}
	\psset{linewidth=1.5,dotsep=1}
	\psset{dotsize=0.7 2.5,dotscale=1 1,fillcolor=black}
	\psset{arrowsize=1 2,arrowlength=1,arrowinset=0.25}
	\begin{pspicture}(2,11)(34,26)
		\psdots[](34,11)(24,26)(4,26)
		\psline[linewidth=1](34,11)(24,26)(4,26)
		\psline[linewidth=1,linestyle=dotted](4,26)(34,11)
		\psline[linestyle=none]{->}(24,26)(13,26)
		\psline[linestyle=none]{->}(34,11)(28,20)
		\psline[linestyle=none]{->}(4,26)(20,18)
\end{pspicture}
	}
}

\newcommand{\xxtow}{\raisebox{-5pt}{
	\psset{unit=0.3 mm}
	\psset{linewidth=1.5,dotsep=1}
	\psset{dotsize=0.7 2.5,dotscale=1 1,fillcolor=black}
	\psset{arrowsize=1 2,arrowlength=1,arrowinset=0.25}
	\begin{pspicture}(4,11)(36,26)
		\psdots[](36,26)(16,26)(6,11)
		\psline[linewidth=1](36,26)(16,26)(6,11)
		\psline[linewidth=1,linestyle=dotted](6,11)(36,26)
		\psline[linestyle=none]{->}(16,26)(10,17)
		\psline[linestyle=none]{->}(36,26)(25,26)
		\psline[linestyle=none]{->}(6,11)(20,18)
	\end{pspicture}
	}
}

\newcommand{\vvtox}{\raisebox{-5pt}{
	\psset{unit=0.3 mm}
	\psset{linewidth=1.5,dotsep=1}
	\psset{dotsize=0.7 2.5,dotscale=1 1,fillcolor=black}
	\psset{arrowsize=1 2,arrowlength=1,arrowinset=0.25}
	\begin{pspicture}(4,12)(36,27)
		\psdots[](36,12)(16,12)(6,27)
		\psline[linewidth=1](36,12)(16,12)(6,27)
		\psline[linewidth=1,linestyle=dotted](6,27)(36,12)
		\psline[linestyle=none]{->}(6,27)(13,17)
		\psline[linestyle=none]{->}(16,12)(26,12)
		\psline[linestyle=none]{->}(36,12)(20,20)
	\end{pspicture}
	}
}

\newcommand{\xtoww}{\raisebox{-5pt}{
	\psset{unit=0.3 mm}
	\psset{linewidth=1.5,dotsep=1}
	\psset{dotsize=0.7 2.5,dotscale=1 1,fillcolor=black}
	\psset{arrowsize=1 2,arrowlength=1,arrowinset=0.25}
	\begin{pspicture}(3,12)(35,27)
		\psdots[](35,27)(25,12)(5,12)
		\psline[linewidth=1](35,27)(25,12)(5,12)
		\psline[linewidth=1,linestyle=dotted](5,12)(35,27)
		\psline[linestyle=none]{->}(5,12)(15,12)
		\psline[linestyle=none]{->}(25,12)(30,19)
		\psline[linestyle=none]{->}(35,27)(20,19)
	\end{pspicture}
	}
}

\newcommand{\uptri}{\raisebox{-5pt}{
	\psset{unit=0.3 mm}
	\psset{linewidth=1.5,dotsep=1}
	\psset{dotsize=0.7 2.5,dotscale=1 1,fillcolor=black}
	\psset{arrowsize=1 2,arrowlength=1,arrowinset=0.25}
	\begin{pspicture}(3,4)(25,21)
		\psdots[](5,4)(25,4)(15,21)
		\pspolygon[linewidth=1](5,4)(25,4)(15,21)(5,4)
		\psline[linestyle=none]{->}(25,4)(20,13)
		\psline[linestyle=none]{->}(15,21)(9,11)
		\psline[linestyle=none]{->}(5,4)(16,4)
	\end{pspicture}
	}
}

\newcommand{\dntri}{\raisebox{-5pt}{
	\psset{unit=0.3 mm}
	\psset{linewidth=1.5,dotsep=1}
	\psset{dotsize=0.7 2.5,dotscale=1 1,fillcolor=black}
	\psset{arrowsize=1 2,arrowlength=1,arrowinset=0.25}
	\begin{pspicture}(3,5)(25,21)
		\psdots[](25,21)(5,21)(15,5)
		\pspolygon[linewidth=1](25,21)(5,21)(15,5)(25,21)
		\psline[linestyle=none]{->}(5,21)(10,12.5)
		\psline[linestyle=none]{->}(15,5)(21,14.4)
		\psline[linestyle=none]{->}(25,21)(14,21)
	\end{pspicture}
	}
}

\newcommand{\yytox}{\raisebox{-5pt}{
	\psset{unit=0.3 mm}
	\psset{linewidth=1.5,dotsep=1}
	\psset{dotsize=0.7 2.5,dotscale=1 1,fillcolor=black}
	\psset{arrowsize=1 2,arrowlength=1,arrowinset=0.25}
	\begin{pspicture}(-2,-2)(17,17)
		\psdots[](0,15)(15,15)(15,0)
		\psline[linewidth=1](0,15)(15,15)(15,0)
		\psline[linewidth=1,linestyle=dotted](15,0)(0,15)
		\psline[linestyle=none]{<-}(4,15)(7,15)
		\psline[linestyle=none]{<-}(15,10)(15,7)
		\psline[linestyle=none]{<-}(9,6)(6,9)
	\end{pspicture}
	}
}

\newcommand{\xxtoyy}{\raisebox{-5pt}{
	\psset{unit=0.3 mm}
	\psset{linewidth=1.5,dotsep=1}
	\psset{dotsize=0.7 2.5,dotscale=1 1,fillcolor=black}
	\psset{arrowsize=1 2,arrowlength=1,arrowinset=0.25}
	\begin{pspicture}(-2,-2)(17,17)
		\psdots[](15,15)(15,0)(0,0)
		\psline[linewidth=1](15,15)(15,0)(0,0)
		\psline[linewidth=1,linestyle=dotted](15,15)(0,0)
		\psline[linestyle=none]{<-}(15,10)(15,7)
		\psline[linestyle=none]{<-}(10,0)(7,0)
		\psline[linestyle=none]{<-}(6,6)(9,9)
	\end{pspicture}
	}
}

\newcommand{\xtoy}{\raisebox{-5pt}{
	\psset{unit=0.3 mm}
	\psset{linewidth=1.5,dotsep=1}
	\psset{dotsize=0.7 2.5,dotscale=1 1,fillcolor=black}
	\psset{arrowsize=1 2,arrowlength=1,arrowinset=0.25}
	\begin{pspicture}(-2,-2)(17,17)
		\psdots[](0,0)(0,15)(15,15)
		\psline[linewidth=1](0,0)(0,15)(15,15)
		\psline[linewidth=1,linestyle=dotted](15,15)(0,0)
		\psline[linestyle=none]{<-}(0,5)(0,8)
		\psline[linestyle=none]{<-}(5,15)(8,15)
		\psline[linestyle=none]{<-}(9,9)(6,6)
	\end{pspicture}
	}
}

\newcommand{\ytoxx}{\raisebox{-5pt}{
	\psset{unit=0.3 mm}
	\psset{linewidth=1.5,dotsep=1}
	\psset{dotsize=0.7 2.5,dotscale=1 1,fillcolor=black}
	\psset{arrowsize=1 2,arrowlength=1,arrowinset=0.25}
	\begin{pspicture}(-2,-2)(17,17)
		\psdots[](15,0)(0,0)(0,15)
		\psline[linewidth=1](15,0)(0,0)(0,15)
		\psline[linewidth=1,linestyle=dotted](0,15)(15,0)
		\psline[linestyle=none]{<-}(10,0)(7,0)
		\psline[linestyle=none]{<-}(0,5)(0,8)
		\psline[linestyle=none]{<-}(6,9)(9,6)
	\end{pspicture}
	}
}

\fi

\ifx\usetikzsymbols\checktikzsymbols

\tikzstyle arrow=[thick,<-]
\tikzstyle shortarrow=[thick,->,shorten <= 9, shorten >= 6]
\tikzstyle longarrow=[thick,->,shorten <= 15, shorten >= 10]
\tikzstyle nn=[thick]
\tikzstyle nnn=[thick,densely dotted]

\newcommand{\wwtov}{\raisebox{-10pt}{
	\begin{tikzpicture}[x=0.3mm, y=0.3mm]
		\filldraw[] (16,4) circle (2);
		\filldraw[] (26,19) circle (2);
		\filldraw[] (16,34) circle (2);
		\draw[nn](16,4) -- (26,19) -- (16,34);
		\draw[nnn](16,34) -- (16,4);
		\draw[shortarrow] (16,4) -- (26,19);
		\draw[shortarrow] (26,19) -- (16,34);
		\draw[longarrow](16,34) -- (16,4);
	\end{tikzpicture}
	}
}

\newcommand{\wtovv}{\raisebox{-10pt}{
	\begin{tikzpicture}[x=0.3mm, y=0.3mm]
		\filldraw[] (24,4) circle (2);
		\filldraw[] (14,19) circle (2);
		\filldraw[] (24,34) circle (2);
		\draw[nn](24,34) -- (14,19) -- (24,4);
		\draw[nnn](24,4) -- (24,34);
		\draw[shortarrow](24,34) -- (14,19);
		\draw[shortarrow](14,19) -- (24,4);
		\draw[longarrow](24,4) -- (24,34);
	\end{tikzpicture}
	}
}

\newcommand{\vtoxx}{\raisebox{-5pt}{
	\begin{tikzpicture}[x=0.3mm, y=0.3mm]
		\filldraw[] (34,11) circle (2);
		\filldraw[] (24,26) circle (2);
		\filldraw[] (4,26) circle (2);
		\draw[nn](34,11) -- (24,26) -- (4,26);
		\draw[nnn](4,26) -- (34,11);
		\draw[shortarrow](34,11) -- (24,26);
		\draw[shortarrow](24,26) -- (4,26);
		\draw[longarrow](4,26) -- (34,11);
\end{tikzpicture}
	}
}

\newcommand{\xxtow}{\raisebox{-5pt}{
	\begin{tikzpicture}[x=0.3mm, y=0.3mm]
		\filldraw[] (36,26) circle (2);
		\filldraw[] (16,26) circle (2);
		\filldraw[] (6,11) circle (2);
		\draw[nn](36,26) -- (16,26) -- (6,11);
		\draw[nnn](6,11) -- (36,26);
		\draw[shortarrow](36,26) -- (16,26);
		\draw[shortarrow](16,26) -- (6,11);
		\draw[longarrow](6,11) -- (36,26);
	\end{tikzpicture}
	}
}

\newcommand{\vvtox}{\raisebox{-5pt}{
	\begin{tikzpicture}[x=0.3mm, y=0.3mm]
		\filldraw[] (36,12) circle (2);
		\filldraw[] (16,12) circle (2);
		\filldraw[] (6,27) circle (2);
		\draw[nn](6,27) -- (16,12) -- (36,12);
		\draw[nnn](36,12) -- (6,27);
		\draw[shortarrow](6,27) -- (16,12);
		\draw[shortarrow](16,12) -- (36,12);
		\draw[longarrow](36,12) -- (6,27);
	\end{tikzpicture}
	}
}

\newcommand{\xtoww}{\raisebox{-5pt}{
	\begin{tikzpicture}[x=0.3mm, y=0.3mm]
		\filldraw[] (35,27) circle (2);
		\filldraw[] (25,12) circle (2);
		\filldraw[] (5,12) circle (2);
		\draw[nn](5,12) -- (25,12) -- (35,27);
		\draw[nnn](35,27) -- (5,12);
		\draw[shortarrow](5,12) -- (25,12);
		\draw[shortarrow](25,12) -- (35,27);
		\draw[longarrow](35,27) -- (5,12);
	\end{tikzpicture}
	}
}

\newcommand{\uptri}{\raisebox{-5pt}{
	\begin{tikzpicture}[x=0.3mm, y=0.3mm]
		\filldraw[] (5,4) circle (2);
		\filldraw[] (25,4) circle (2);
		\filldraw[] (15,21) circle (2);
		\draw[nn](5,4) -- (25,4) -- (15,21) -- cycle;
		\draw[shortarrow](5,4) -- (25,4);
		\draw[shortarrow](25,4) -- (15,21);
		\draw[shortarrow](15,21) -- (5,4);
	\end{tikzpicture}
	}
}

\newcommand{\dntri}{\raisebox{-5pt}{
	\begin{tikzpicture}[x=0.3mm, y=0.3mm]
		\filldraw[](25,21) circle (2);
		\filldraw[](5,21) circle (2);
		\filldraw[](15,5) circle (2);
		\draw[nn] (25,21) -- (5,21) -- (15,5) -- cycle;
		\draw[shortarrow] (25,21) -- (5,21);
		\draw[shortarrow] (5,21) -- (15,5);
		\draw[shortarrow] (15,5) -- (25,21);
	\end{tikzpicture}
	}
}

\newcommand{\yytox}{\raisebox{-5pt}{
	\begin{tikzpicture}[x=0.3mm, y=0.3mm]
		\filldraw[](0,15) circle (2);
		\filldraw[](15,15) circle (2);
		\filldraw[](15,0) circle (2);
		\draw[nn](0,15) -- (15,15) -- (15,0) ;
		\draw[nnn](15,0) -- (0,15) ;
		\draw[arrow](5,15) -- (8,15) ;
		\draw[arrow](15,10) -- (15,7) ;
		\draw[arrow](9,6) -- (6,9) ;
	\end{tikzpicture}
	}
}

\newcommand{\xxtoyy}{\raisebox{-5pt}{
	\begin{tikzpicture}[x=0.3mm, y=0.3mm]
		\filldraw[](15,15) circle (2);
		\filldraw[](15,0) circle (2);
		\filldraw[](0,0) circle (2);
		\draw[nn](15,15) -- (15,0) -- (0,0) ;
		\draw[nnn](15,15) -- (0,0) ;
		\draw[arrow](15,10) -- (15,7) ;
		\draw[arrow](10,0) -- (7,0) ;
		\draw[arrow](6,6) -- (9,9) ;
	\end{tikzpicture}
	}
}

\newcommand{\xtoy}{\raisebox{-5pt}{
	\begin{tikzpicture}[x=0.3mm, y=0.3mm]
		\filldraw[](0,0) circle (2);
		\filldraw[](0,15) circle (2);
		\filldraw[](15,15) circle (2);
		\draw[nn](0,0) -- (0,15) -- (15,15) ;
		\draw[nnn](15,15) -- (0,0) ;
		\draw[arrow](0,5) -- (0,8) ;
		\draw[arrow](5,15) -- (8,15) ;
		\draw[arrow](9,9) -- (6,6) ;
	\end{tikzpicture}
	}
}

\newcommand{\ytoxx}{\raisebox{-5pt}{
	\begin{tikzpicture}[x=0.3mm, y=0.3mm]
		\filldraw[](15,0) circle (2);
		\filldraw[](0,0) circle (2);
		\filldraw[](0,15) circle (2);
		\draw[nn](15,0) -- (0,0) -- (0,15) ;
		\draw[nnn](0,15) -- (15,0) ;
		\draw[arrow](10,0) -- (7,0) ;
		\draw[arrow](0,5) -- (0,8) ;
		\draw[arrow](6,9) -- (9,6) ;
	\end{tikzpicture}
	}
}

\fi

\ifx\usetikzsymbols\undefined
\ifx\usepstsymbols\undefined

\newcommand{\wwtov}{\raisebox{-10pt}{\includegraphics{ww2v.eps}}}
\newcommand{\wtovv}{\raisebox{-10pt}{\includegraphics{w2vv.eps}}}
\newcommand{\vtoxx}{\raisebox{-5pt}{\includegraphics{v2xx.eps}}}
\newcommand{\xxtow}{\raisebox{-5pt}{\includegraphics{xx2w.eps}}}
\newcommand{\vvtox}{\raisebox{-5pt}{\includegraphics{vv2x.eps}}}
\newcommand{\xtoww}{\raisebox{-5pt}{\includegraphics{x2ww.eps}}}
\newcommand{\uptri}{\raisebox{-5pt}{\includegraphics{uptriangle.eps}}}
\newcommand{\dntri}{\raisebox{-5pt}{\includegraphics{dntriangle.eps}}}

\fi
\fi

\ifx\filesasinput\undefined
\begin{document}
Here is some texts with the \yytox inline graphics. Here \xxtoyy, \ytoxx, \xtoy are some more. And here \wwtov, \wtovv, \vtoxx, \xxtow, \vvtox, \xtoww, \uptri, \dntri are even more.
\end{document}
\fi

\usepackage[dvips]{hyperref}

\begin{document}

\title{Proposal for detecting spin-chirality terms in Mott insulators via resonant inelastic x-ray scattering}

\author{Wing-Ho Ko}
\affiliation{Kavli Institute for Theoretical Physics, University of California, Santa Barbara, Santa Barbara, California 93106, USA}
\author{Patrick A. Lee}
\affiliation{Department of Physics, Massachusetts Institute of Technology, Cambridge, Massachusetts 02139, USA}

\date{July 27, 2011}

\begin{abstract} 
We consider the question of whether resonant inelastic x-ray scattering (RIXS) can be used to detect many-body excitations that are coupled to the spin-chirality terms $\SSS{i}{j}{k}$ in a Mott insulator. We find that while the spin-chirality terms are in general absent in the usual experimental setups of RIXS, there are prospects of realizing such terms if one considers instead the scattering near a \emph{pre-edge}. We then perform detailed analyses for the square and the kagome lattices, and brief analyses for the triangular and the honeycomb lattices, in which we show that the spin-chirality terms are indeed present in all the above lattices, but that they occur at a higher order in our expansion for the kagome and the honeycomb lattices.  The merit of using RIXS in addition to Raman spectroscopy to detect excitations that are coupled to the spin-chirality terms is also briefly discussed in the context of the emergent gauge boson in the $U(1)$ Dirac spin liquid.
\end{abstract}

\maketitle 

\section{Introduction} \label{sect:intro} 

With its ability to probe generic many-body excitations, Raman spectroscopy has become an important tool for understanding strongly correlated electronic systems, including the Mott insulators.\cite{Devereaux:RMP:2007} Unfortunately, in Raman spectroscopy the photon momenta are generally negligible when compared with the inverse lattice scale, thus making it essentially a zero-momentum-transfer probe and limiting its usefulness in certain cases.


In the specific context of the $U(1)$ Dirac spin liquid state in the spin-1/2 kagome lattice, in which the low-energy effective theory is described by chargeless spin-1/2 fermions (spinons) coupled to an emergent $U(1)$ gauge field,\cite{Hermele:PRB:2008} we previously proposed Raman spectroscopy as a way to detect the spinon continuum and the fluctuations of the emergent gauge field.\cite{Ko:PRB:2010} The prospect of detecting the emergent gauge field in experiments is particularly significant, given the role such gauge fields have played in theories of quantum spin liquids. Unfortunately, our calculations also reveal that the signal coming from the emergent gauge field is suppressed by a factor of $q^2$, where $q$ is the momentum transferred to the system. Indeed, Raman spectroscopy on herbertsmithite ZnCu$_3$(OH)$_6$Cl$_2$, a possible material realization of the $U(1)$ Dirac spin liquid state, has found a broad continuum in the spectrum that could be attributed to the spinon continuum, but has shown no signs of the emergent gauge field.\cite{Wulferding:PRB:2010}

Given the large momentum carried by x-ray, one would imagine that resonant inelastic x-ray scattering (RIXS)\cite{Kotani:RMP:2001} may provide a better prospect of detecting the emergent gauge field. However, in our derivation, the detection of the emergent gauge field in Raman spectroscopy depends crucially on the coupling of the external photons to the spin-chirality terms $\vv{S}_i \cdot (\vv{S}_j \times \vv{S}_k)$ in the system, which in turn relies crucially on the link between the photon polarizations and the direction of the virtual electron hops in the lattice induced by the virtual absorption and emission of the photons. This link is in general absent in the current theoretical discussions and experimental setups of RIXS, in which the virtual absorption and emission of the photons are accompanied by \emph{intra-site} electron hops. 

In this paper, we propose performing RIXS near a \emph{pre-edge}, in which case the usual virtual processes with intra-site photon-induced electron hops are suppressed, thus allowing virtual processes with \emph{inter-site} photon-induced electron hops to manifest. Indeed, inter-site dipolar contributions have previously been identified in the absorption and Auger spectrum of\cite{Uozumi:EPL:1992, Danger:PRL:2002, Uozumi:JESRP:2004} TiO$_2$ and\cite{Shukla:PPL:2006, Kotani:PRB:2008} La$_2$CuO$_4$. To analyze the contributions by such inter-site processes to the RIXS signals, we modify the Shastry--Shraiman formalism\cite{Shastry:PRL:1990,Shastry:IJMPB:1991} used in deriving the corresponding results in Raman spectroscopy, and show that the spin-chirality terms indeed appear in both the square lattice (cuprate) and the kagome lattice (herbertsmithite), but that the first appearance of such terms occurs at a higher order in the latter case.

This paper is organized as follows: in Sec.~\ref{sect:Review} the Shastry--Shraiman formalism in Raman spectroscopy is reviewed to set the stage for Sec.~\ref{sect:RIXS}, in which the modifications to this formalism to the case of RIXS are discussed and illustrated. In Sec.~\ref{sect:SSS} the possibility of detecting the spin-chirality terms in RIXS is considered in this modified formalism and our new proposal is presented, followed by detailed analyses for the square and the kagome lattices, as well as brief discussions for the triangular and the honeycomb lattices. Further discussions ensue in Sec.~\ref{sect:discussions}, in which our motivating example of the $U(1)$ Dirac spin liquid is considered again to put our proposal into perspective.

\section{Review of the Shastry--Shraiman formalism in Raman Spectroscopy} \label{sect:Review}

In the Shastry--Shraiman formalism, the electron-photon interaction $H_C$ is treated as a time-dependent perturbation on the time-independent Hamiltonian $H_{\txt{ind}}$. The latter consists of the Hubbard Hamiltonian $H_{\txt{Hb}} = \sum_{ij,\sigma} t_{ij} c^{\dg}_{i\sigma} c_{j\sigma} + U \sum_i n_{i\up} n_{i\dn}$ and the free photon Hamiltonian $H_{\gamma} = \sum_{\vv{q},\alpha} \omega_{\vv{q}} a^{\alpha\dg}_{\vv{q}} a^\alpha_{\vv{q}}$ (here $\alpha$ labels the photon polarizations). Applying Fermi's golden rule, the transition rate $W_{fi}$ from an initial state $\ket{i}$ to a final state $\ket{f}$ is given by:
\begin{equation} \label{eq:Fermi}
W_{fi} = 2\pi |\bra{f} T \ket{i}|^2 \delta(\E_f - \E_i) \punct{,}
\end{equation}
where $T$ is the scattering $T$-matrix. Keeping only terms that are second order in the photon operators $a^{\alpha}_{\vv{q}}$, the $T$-matrix can be decomposed as $T = T_{\txt{R}} + T_{\txt{NR}}$, with $T_{\txt{R}}$ the resonant part and $T_{\txt{NR}}$ the non-resonant part, of which only the former is important in a Mott insulator. Next, the energy denominator that appears in $T_{\txt{R}}$ is further expanded by treating the hopping part $H_t = \sum_{ij,\sigma} t_{ij} c^{\dg}_{i\sigma} c_{j\sigma}$ of the Hubbard Hamiltonian as a perturbation on the remaining terms in $H_{\txt{ind}}$. To be more precise,
\begin{align}
T_\txt{R} & = H^{(1)}_{C} \frac{1}{\E_i - (H_{\txt{Hb}} + H_{\gamma}) + i \eta} H^{(1)}_{C} \label{eq:TR_def} \\
	& = H^{(1)}_{C} \frac{1}{\E_i - H_U - H_\gamma + i\eta} \notag \\
	& \quad \times \sum_{n=0}^{\infty} \left( H_t \frac{1}{\E_i - H_U - H_\gamma + i\eta} \right)^n H^{(1)}_{C} \punct{,} 	\label{eq:TR_expand}
\end{align}
where $\E_i$ is the initial energy of the unperturbed system, $H^{(1)}_{C}$ consists of the terms in $H_C$ that are first order in the photon operators $a^{\alpha}_{\vv{q}}$, and $H_U = U \sum_i n_{i\up} n_{i\dn}$ is the on-site Coulomb repulsion in the Hubbard Hamiltonian. In the context of Raman scattering, $H^{(1)}_C \sim \sum_{ij,\sigma} i t_{ij} c^\dg_{i\sigma} c_{j\sigma} (g_i \vv{e}_\alpha a_{\vv{k}}^\alpha e^{i\vv{k}\cdot\vv{x}} + g_f \overline{\vv{e}}_\beta a_{\vv{q}}^\beta e^{-i\vv{q}\cdot\vv{x}})$, where $g_i$ ($g_f$) denotes the appropriate coupling constants between the electron and the incoming (outgoing) photon, $\vv{x} = (\vv{x}_i + \vv{x}_j)/2$ is the mid-point between site $i$ and $j$, and $\vv{e}_\alpha$ ($\vv{e}_\beta$) and $\vv{k}$ ($\vv{q}$) denote the polarization vector and momentum of the incoming (outgoing) photon, respectively.

\newcommand{\fnOne}{One caveat is that $\E_i$ in principle depends on the initial state, which is assumed to be an eigenstate of the time-independent Hamiltonian $H_{\txt{ind}}$. An arbitrary spin state in the lattice basis will in general not be an eigenstate of $H_{\txt{ind}}$ and technically $\E_i$ itself has to be treated perturbatively. This subtlety can be neglected when working in the zeroth order in $t/U$. Moreover, at the end of the day the $T$-matrix will be evaluated with respect to some many-body states that are the approximate ground states or low-lying excited states of $H_{\txt{ind}}$ within specific models, in which case it is legitimate to treat $\E_i$ as a $c$-number, to be absorbed into the definition of the resonant frequency.}

In a Mott insulator, $T_{\txt{R}}$ connects between two spin states (i.e., states with zero double occupancy) and hence can in principle be expressed in terms of spin operators. In the Shastry--Shraiman formalism, this is achieved by inserting a complete set of states (in the lattice occupation basis with a fixed spin quantization axis) in between the operators in Eq.~(\ref{eq:TR_expand}), under which the energy denominators $(\E_i - H_U - H_\gamma + i\eta)^{-1}$ become $c$-numbers.\footnote{\fnOne} Moreover, once an initial spin state (in that same basis) is specified and a particular choice of individual term is chosen for each $H_t$ and $H^{(1)}_C$ in Eq.~(\ref{eq:TR_expand}), the intermediate states are uniquely determined and thus can be trivially resummed. Hence the matrix elements of $T_{\txt{R}}$ with respect to spin states can be expressed as a sum of chains of electronic operators. These chains of electronic operators can be visualized as virtual processes in which electrons hop around the lattice and can be converted into spin operators by the identities $\tchi_{\sigma\sigma'} \equiv c^\dg_{\sigma'} c_{\sigma} = \frac{1}{2} \delta_{\sigma \sigma'} + \vv{S} \cdot \vvsym{\tau}_{\sigma \sigma'}$ and $\chi_{\sigma\sigma'} \equiv c_{\sigma} c^\dg_{\sigma'} = \frac{1}{2} \delta_{\sigma \sigma'} - \vv{S} \cdot \vvsym{\tau}_{\sigma \sigma'}$. For $t \ll U$ and near resonance (i.e., $\omega_i \approx U$, in which $\omega_i$ is the energy of the incoming photon), the contributions to $T_{\txt{R}}$ are dominated by virtual processes in which all intermediate states have exactly one hole and one doubly occupied site (a.k.a.\@ doublon). In such case $(\E_i - H_U - H_\gamma + i\eta) \approx (\omega_i - U)$, and the matrix elements of $T_{\txt{R}}$ can thus be organized as an expansion in $t/(\omega_i - U)$. 

\begin{figure}
\begin{center}
\subfigure[\label{fig:Raman_1}]{\includegraphics[scale=1]{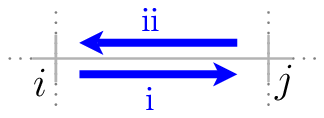}} \\
\subfigure[\label{fig:Raman_2}]{\includegraphics[scale=1]{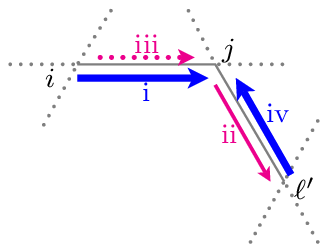}} \quad
\subfigure[\label{fig:Raman_3}]{\includegraphics[scale=1]{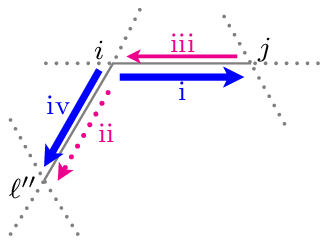}}
\caption{\label{fig:Raman} \colorfig Virtual processes that contribute to the resonant Raman scattering $T$-matrix $T_{\txt{R}}$. The process depicted in (a) contributes to the Fleury--Loudon term while the ones depicted in (b) and (c) contribute (among others) to the spin-chirality terms at the leading order. Here and henceforth thick (blue) arrows denote electron hops that are accompanied by virtual absorptions or emissions of photons, thin (magenta) unbroken arrows denote movements of electrons in non-photon-induced internal hops, and thin (magenta) broken arrows denote movements of holes in non-photon-induced internal hops. The order of hops is indicated by lowercase roman letters next to the corresponding arrows.}
\end{center}
\end{figure}

To the lowest nontrivial order in $t/(\omega_i - U)$, the $T$-matrix obtained in the Shastry--Shraiman formalism reproduces the Fleury--Loudon Hamiltonian $H_{\txt{FL}} = \sum_{\vv{r},\vv{r}'} \frac{2 t_{\vv{r}\vv{r}'}^2}{U-\omega_i}  (\vv{e}_i \cdot \vvsym{\mu}) (\overline{\vv{e}}_f \cdot \vvsym{\mu}) (1/4 - \vv{S}_{\vv{r}} \cdot \vv{S}_{\vv{r}'} ) $, \cite{Fleury:PR:1968} and is contributed by virtual processes of the form shown in Fig.~\ref{fig:Raman_1}. At the $t^4/(\omega_i - U)^3$ order, individual processes that contribute to the spin-chirality terms, such as the ones shown in Figs.~\ref{fig:Raman_2} and \ref{fig:Raman_3}, start to appear. However, in the case where only nearest-neighbor hoppings are included, the sum of their contributions is found to vanish in the square and the triangular lattices, while it remains nonzero in the honeycomb and the kagome lattices.\cite{Ko:PRB:2010} It is worth noting that the spin-chirality terms appear exclusively in the $(\efei{x}{y} -\efei{y}{x})$ polarization channel in the scattering $T$-matrix.

\section{Modification of the Shastry--Shraiman formalism to RIXS} \label{sect:RIXS}

Since both Raman scattering and RIXS are resonant two-photon processes, it should be possible to modify the Shastry--Shraiman formalism to the case of RIXS. Indeed, similar expansion of $T_R$ had been made by van den Brink and van Veenendaal.\cite{Brink:EPL:2006} However, in that work the denominator in Eq.~(\ref{eq:TR_def}) was expanded by treating the (appropriately modified, see below) Hubbard Hamiltonian as a perturbation on the free photon and the atomic (see below) Hamiltonians. In practice, the Hubbard Hamiltonian is then expanded in the usual way one derives the Heisenberg Hamiltonian.\cite{Brink:EPL:2007, Forte:PRB:2008} For the present work, we follow instead the spirit of the Shastry--Shraiman formalism and take the terms in $H_{\txt{ind}}$ in which site indices change as perturbations on the free photon and the on-site terms in $H_{\txt{ind}}$. Since the main purpose of this paper is to identify virtual processes that may give rise to the spin-chirality terms, of which the two expansion schemes agree except that the prefactors coming from the energy denominators are organized differently, we shall not dwell on the relative merit of these two expansions, which may depend on one's identification of the resonant energy. Instead, we simply state here the necessary modifications to the Shastry--Shraiman formalism in the case of RIXS.

First, $H^{(1)}_C$ now corresponds to virtual transitions in which an electron hops from a core state to a valence state while a photon is absorbed, or in which an electron hops from a valence state to the core state while a photon is emitted. Hence we may write $H^{(1)}_C = \sum_{c,v} i g_i \vv{e}_{\alpha} c^\dg_{v} J^{\alpha}_{vc} c_{c} a^{\alpha}_{\vv{k}} e^{i\vv{k} \cdot \vv{r}} + i g_f \overline{\vv{e}}_{\beta} c^\dg_{c} (J^\dg)^{\beta}_{cv} c_{v} a^{\beta}_{\vv{q}} e^{i\vv{q} \cdot \vv{r}'}$, where the subscript $c$ ($v$) labels a core (valence) state and $J$ is a (possibly polarization-dependent) matrix that accounts for the matrix elements of the atomic transitions.\cite{Groot:PRB:1998} As usual it is necessary to include only the core and valence states that are near resonance in $H^{(1)}_C$.

Second, $H_{\txt{ind}}$ must now include extra terms that account for the single-particle energies of the core states and possibly of the high-energy valence states, as well as their interactions with the low-energy valence states. Schematically, we can write $H_{\txt{ind}} = H_{\gamma} + H_{\txt{atomic}} + H'_{\txt{Hb}}$, in which the atomic Hamiltonian  $H_{\txt{atomic}}$ accounts for the energy difference between the core state and the valence state excited from it, while the modified Hubbard Hamiltonian $H'_{\txt{Hb}}$ accounts for the interactions of the valence electrons among themselves, with the lattice potential, and with the core hole. In particular, since $H_U$ captures only the low-energy effective Coulomb repulsion among the low-energy valence electrons, in principle it is necessary to include in $H'_{\txt{Hb}}$ generic Coulomb interactions $u_{\alpha\beta\gamma\delta} c^{\dg}_{\alpha} c^{\dg}_{\beta} c_{\gamma} c_{\delta}$ in which at least one of the electron operators corresponds to a core or high-energy valence state. However, on physical ground it may be argued that the dominant effect of the core or high-energy valence state on the low-energy valence states would be modifications to the hopping parameters and the on-site potentials, i.e., $u_{\alpha\beta\gamma\delta} c^{\dg}_{\alpha} c^{\dg}_{\beta} c_{\gamma} c_{\delta} \sim c^\dg_{c} c_{c} V_{c, ij} c^\dg_{i} c_{j}$ and $c^\dg_{e} c_{e} V_{e, ij} c^\dg_{i} c_{j}$, where the subscript $c$ ($e$) labels a core (high-energy valence) state while $i,j$ are site labels of low-energy valence states. In practice, the effect of the core hole may be well captured by an on-site potential $U_c$ localized at the site where the core hole is present.\cite{Forte:PRB:2008} 

Third, unlike in Raman scattering, the core hole in RIXS has a very short lifetime and can decay via Auger processes. This introduces an uncertainty to the core-hole energy, which can be captured by replacing the infinitesimal $\eta$ in Eqs.~(\ref{eq:TR_def})~and~(\ref{eq:TR_expand}) by a finite energy broadening $\Gamma$.

\begin{figure}
\includegraphics[scale=0.8]{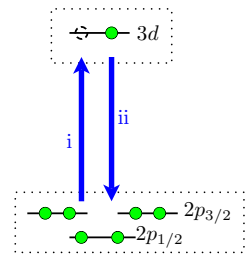}
\caption{\label{fig:2p3d} \colorfig The lowest-order virtual processes that contribute to the one-magnon excitation in $2p \rightarrow 3d$ RIXS.}
\end{figure}

To illustrate the modified Shastry--Shraiman formalism, we briefly outline the virtual processes that contribute to the single-magnon excitation in $2p \rightarrow 3d$ RIXS and to the two-magnon excitation in $1s \rightarrow 4p$ RIXS in cuprates, both of which have previously been proposed\cite{Groot:PRB:1998, Forte:PRB:2008} and experimentally studied.\cite{Duda:Uppsala:1996,Hill:PRL:2008} 

For the $2p \rightarrow 3d$ RIXS, the lowest-order processes that contribute to the single-magnon excitations are purely atomic in nature and involve simply the photon-induced virtual transitions of a $2p$ core electron to and from the $3d$ valence states (see Fig.~\ref{fig:2p3d} for illustration). Since the $2p$ states are spin-orbit coupled, spin is not a good quantum number, and a spin flip of the $3d$ electrons can occur, as long as it is accompanied by an appropriate change in the photon polarization. In the modified Shastry--Shraiman formalism, the chain of electron operators that is associated with the process depicted in Fig.~\ref{fig:2p3d} is:
\begin{align}
T_{\txt{1-magnon}} & \propto \sum_{p,p',\sigma,\sigma'} 
	(c^\dg_{p} (J^\dg)^{\beta}_{p\sigma} c_\sigma) (c^\dg_{\sigma'} J^\alpha_{\sigma' p'} c_{p'}) \notag \\
	& = \tr \{ M^{\alpha\beta} \chi \} \notag \\
	& = m^{\alpha\beta}_0 - 2 \vv{m}^{\alpha\beta} \cdot \vv{S} \punct{,} \label{eq:2p3d}
\end{align}
where the subscript $p$ labels the $2p$ states while the subscript $\sigma$ labels the spin of the valence $3d$ states. Also, $\chi_{\sigma\sigma'} \equiv c_{\sigma} c^\dg_{\sigma'} = \frac{1}{2} \delta_{\sigma\sigma'} - \vv{S} \cdot \vvsym{\tau}_{\sigma\sigma'}$ as before while $M^{\alpha\beta}_{\sigma' \sigma} = m^{\alpha\beta}_0 \delta_{\sigma'\sigma} + \vv{m}^{\alpha\beta} \cdot \vvsym{\tau}_{\sigma' \sigma} = \sum_{p} J^{\alpha}_{\sigma' p} (J^\dg)^{\beta}_{p \sigma} $. Note that we have adopted a matrix convention for spin indices in the second line (which will be similarly adopted henceforth). From Eq.~(\ref{eq:2p3d}) it can be seen that the spin-flip term arises from structure of the $2p \rightarrow 3d$ transition matrix elements.

\begin{figure}
\begin{center}
\subfigure[\label{fig:1s4p_1}]{\includegraphics[scale=0.7]{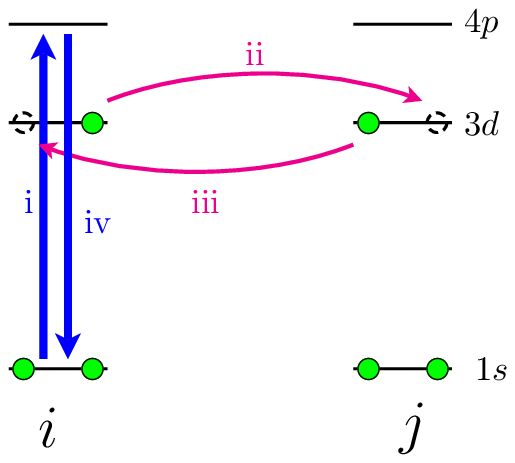}} \qquad
\subfigure[\label{fig:1s4p_2}]{\includegraphics[scale=0.7]{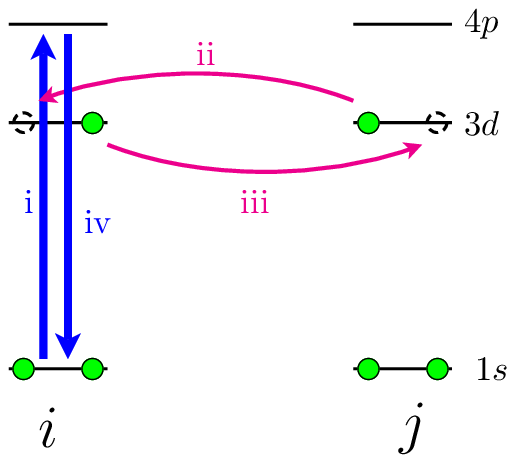}}
\caption{\label{fig:1s4p} \colorfig The lowest-order virtual processes that contribute to the two-magnon excitation in $1s \rightarrow 4p$ RIXS.}
\end{center}
\end{figure}

For the $1s \rightarrow 4p$ RIXS, two-magnon excitation occurs when the core hole and $4p$ valence electron ``shake up'' the low-energy valence electrons and induce a pair exchange. Two such processes are depicted in Fig.~\ref{fig:1s4p}. The chains of electron operators that are associated with the processes depicted in Figs.~\ref{fig:1s4p_1} and \ref{fig:1s4p_2} are, respectively:
\begin{align}
T^{(a)}_{\txt{2-magnon}} & \propto
	(c^\dg_{is} (J^\dg)^{\beta}_{sp} c_{ip}) \cc[t_{ij}]{i}{j} \cc[t_{ji}]{j}{i} (c^\dg_{ip'} J^\alpha_{p' s'} c_{is'}) \notag \\
& = \tr \{ J^{\alpha} (J^\dg)^{\beta} \} t_{ij}t_{ji} \tr \{ \chi_j \tchi_i \} \notag \\
& = N^{\alpha\beta} t_{ij} t_{ji} \left( \frac{1}{2} - \vv{S}_i \cdot \vv{S}_j  \right) 
	\punct{,} \label{eq:1s4p_1} \\
T^{(b)}_{\txt{2-magnon}} & \propto
	(c^\dg_{is} (J^\dg)^{\beta}_{sp} c_{ip}) \cc[t_{ji}]{j}{i} \cc[t_{ij}]{i}{j} (c^\dg_{ip'} J^\alpha_{p' s'} c_{is'}) \notag \\
& = \tr \{ J^{\alpha} (J^\dg)^{\beta} \} t_{ji} t_{ij} \tr \{ \chi_i \tchi_j \} \notag \\
& = N^{\alpha\beta} t_{ij} t_{ji} \left( \frac{1}{2} - \vv{S}_i \cdot \vv{S}_j  \right)
	\punct{,} \label{eq:1s4p_2}
\end{align}
where $N^{\alpha\beta} \equiv \tr \{ J^{\alpha} (J^\dg)^{\beta} \}$, the subscript $s$ ($p$) labels the $1s$ ($4p$) states, and the electron operators with no orbital labels are assumed to be that of the valence $3d$ states. For brevity here and henceforth we omit the spin indices in virtual hops that involve only the $3d$ electrons, assuming that they are appropriately summed within parentheses. Thus, e.g., $(t_{ij} c^\dg_{i} c_{j}) \equiv \sum_\sigma t_{ij} c^\dg_{i\sigma} c_{j\sigma}$. Similarly, here and henceforth the sums over the spin and (for core hole and high-energy valence electrons) the orbital indices are assumed in the two photon-induced hops, such that, e.g., $(c^\dg_{ip'} J^{\alpha}_{p's'} c_{is'}) \equiv \sum_{p',s'} c^\dg_{ip'} J^{\alpha}_{p's'} c_{is'}$. Note also that the matrix notation in the second line of Eqs.~(\ref{eq:1s4p_1}) and (\ref{eq:1s4p_2}) is now extended to include the orbital indices of the core and the $4p$ electrons.

Under the assumption that the only effect of the $1s$ core hole and the $4p$ valence electron is to introduce an additional potential $U_c$ at the site $i$ of which the atomic transition occurs, the coefficients of $T^{(a)}_{\txt{2-magnon}}$ and $T^{(b)}_{\txt{2-magnon}}$ that come from the energy denominators are, respectively:
\begin{align}
C^{(a)}_{\txt{2-magnon}} & = \left( \frac{1}{\delta\omega + i \Gamma} \right)^2 \frac{1}{\delta\omega -(U + U_c) + i \Gamma} \punct{,} \label{eq:1s4p_1_denom} \\
C^{(b)}_{\txt{2-magnon}} & = \left( \frac{1}{\delta\omega + i \Gamma} \right)^2 \frac{1}{\delta\omega - (U - U_c) + i \Gamma} \punct{.} \label{eq:1s4p_2_denom} 
\end{align}
Here $\delta\omega = \omega_i - (\E_{4p} - \E_{1s})$ is the detuning from the atomic transition. We remark that one must subtract from $C_{\txt{2-magnon}}$ the corresponding coefficients with $U_c = 0$ to obtain the actual contributions of these two processes to the two-magnon \emph{transition}, since the contributions of these two processes with $U_c$ set to $0$, together with other virtual processes in which the intermediate virtual exchange does not involve the site $i$, constitute a part of the $T$-matrix that is proportional to the Heisenberg Hamiltonian and thus is not responsible for any actual transitions.

\begin{figure}
\begin{center}
\subfigure[\label{fig:SSS_fail_1}]{\includegraphics[scale=0.65]{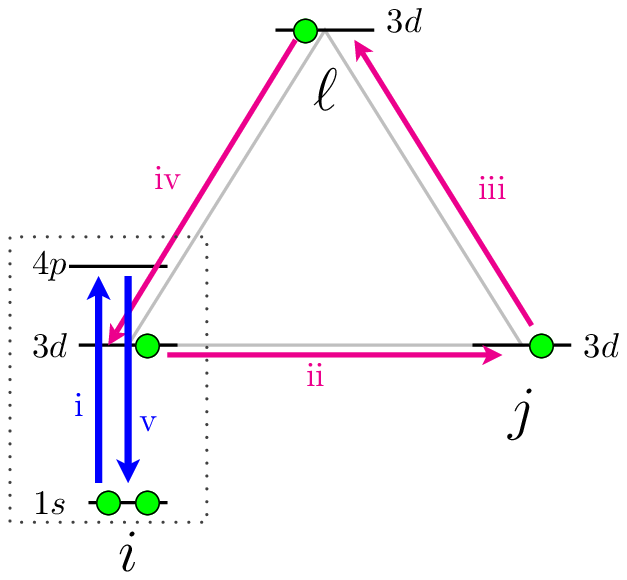}} \quad
\subfigure[\label{fig:SSS_fail_2}]{\includegraphics[scale=0.65]{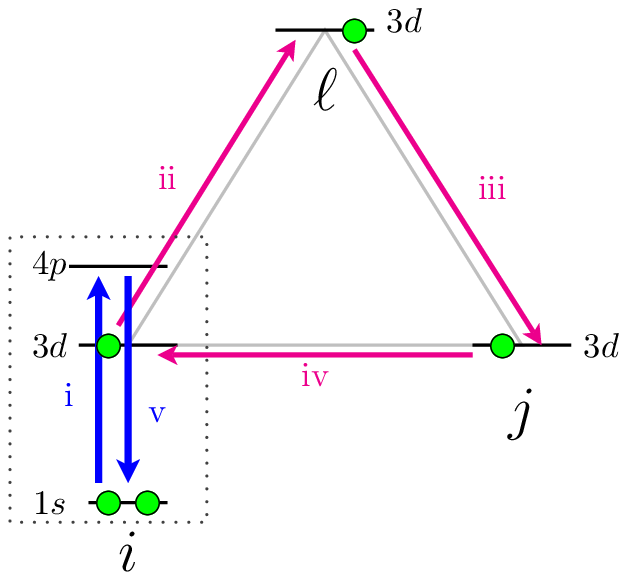}}
\caption{\label{fig:SSS_fail} \colorfig Two processes whose contributions to the spin-chirality terms cancel out each other.}
\end{center}
\end{figure}

\section{Spin-chirality terms in RIXS} \label{sect:SSS}

To obtain the spin-chirality terms, at least three lattice sites must be involved in the virtual processes. For example, one might consider the higher order processes in the $1s \rightarrow 4p$ RIXS shown in Figs.~\ref{fig:SSS_fail_1} and \ref{fig:SSS_fail_2}. From the modified Shastry--Shraiman formalism, the associated chains of electron operators are, respectively:
\begin{align}
T^{(a)}_{\txt{3-sites}} & \propto 
	(c^\dg_{is} (J^\dg)^{\beta}_{sp} c_{ip}) \cc[t_{i\ell}]{i}{\ell} \cc[t_{\ell j}]{\ell}{j}  \cc[t_{ji}]{j}{i} 
	(c^\dg_{i p'} J^\alpha_{p' s'} c_{i s'}) \notag \\
& = \tr \{ J^{\alpha} (J^\dg)^{\beta} \} t_{i\ell}t_{\ell j} t_{ji} 
	\tr \{ \chi_\ell \chi_j \tchi_i \} \notag \\
& = N^{\alpha\beta} t_{i\ell}t_{\ell j} t_{ji} \left(2 i \SSS{\ell}{j}{i} + \ldots\right) \punct{,} \label{eq:SSS_fail_1}  
\end{align}
\begin{align}
T^{(b)}_{\txt{3-sites}} & \propto 
	(c^\dg_{is} (J^\dg)^{\beta}_{sp} c_{ip}) \cc[t_{ij}]{i}{j} \cc[t_{j\ell}]{j}{\ell} \cc[t_{\ell i}]{\ell}{i}
	(c^\dg_{i p'} J^\alpha_{p' s'} c_{i s'}) \notag \\
& = \tr \{ J^{\alpha} (J^\dg)^{\beta} \} t_{ij}t_{j \ell} t_{\ell i} 
	\tr \{ \chi_j \chi_\ell \tchi_i \} \notag \\
& = N^{\alpha\beta} t_{i\ell}t_{\ell j} t_{ji} \left(2 i \SSS{j}{\ell}{i} + \ldots \right) \punct{.} \label{eq:SSS_fail_2}
\end{align}
Moreover, it can be seen that the coefficient coming from the energy denominators is $C_{\txt{3-sites}} = (\delta\omega + i \Gamma)^{-2} (\delta\omega - (U - U_c) + i \Gamma)^{-2}$ for both processes. Hence, while each process by itself contributes to the spin-chirality terms, the sum of their contributions vanishes.\footnote{Note however that the sum of their contributions to the $\vv{S} \cdot \vv{S}$ terms (omitted in the $\ldots$) does not vanish.} Similar calculations show that the cancellation also occurs in the two analogous virtual processes in which electrons hop around the four-site loop in the square lattice. Furthermore, it can be shown that such cancellations also occur for similar virtual processes in the $2p \rightarrow 3d$ RIXS.
 
In Raman scattering, the analogous processes, in which the hops \textit{i} and \textit{v} in Fig.~\ref{fig:SSS_fail} do not exist and which the hops \textit{ii} and \textit{iv} are photon induced, do not cancel out each other. Instead, the anticlockwise loop in Fig.~\ref{fig:SSS_fail_1} contributes to the $\efei{y}{x}$ photon polarization channel while the clockwise loop in Fig.~\ref{fig:SSS_fail_2} contributes to the $\efei{x}{y}$ channel, resulting in a nonvanishing contribution to the $(\efei{x}{y} - \efei{y}{x})$ channel when summed. This suggests that in order for the spin-chirality terms to be realized in the scattering $T$-matrix, it is crucial for the photon polarizations to be coupled with the directions of \emph{inter-site} electron hops---a link that does not appear in the usual RIXS setups.

That said, one should also note that the dipole moment between a core orbital at one site and a valence orbital at one of its nearest-neighbor site is in general nonvanishing. Thus, in principle, a photon from the incident x-ray beam can also induce a core-to-valence excitation across the two sites. Such an inter-site transition is in general suppressed by the reduced wavefunction overlap and is thus masked by the corresponding intra-site transition. Moreover, for hard x-ray the distance between two nearest-neighbor sites may also be equal to many wavelengths of the incident x-ray, which further reduces the transition amplitude for such an inter-site transition at near-horizontal incidence (relative to the two-dimensional lattice plane) as a result of the rapid oscillation of the electric field across the two sites.

However, if the frequency of the incident x-ray is tuned to that of a forbidden atomic transition (e.g., the $1s \rightarrow 3d$ transition in the Cu$^{2+}$ materials), then the near-resonant dipole inter-site transition needs only to compete with a near-resonant \emph{quadruple} intra-site transition and a \emph{detuned} dipole intra-site transition. With adequate luminosity, this may allow the signals from the inter-site transition to manifest in the spectrum. Indeed, contributions from such inter-site transitions have previously been identified in the x-ray absorption and Auger spectroscopy of\cite{Uozumi:EPL:1992, Danger:PRL:2002, Uozumi:JESRP:2004} TiO$_2$ and\cite{Shukla:PPL:2006, Kotani:PRB:2008} La$_2$CuO$_4$.

Moreover, for nearly two-dimensional materials such as cuprates and herbertsmithite, the rapid oscillation of the electric field between two lattice sites can be alleviated by arranging the x-ray to be at near-normal incidence relative to the two-dimensional lattice plane. Generally, by tuning the angle of incidence, the electric field across two nearest-neighbor sites can be made relatively uniform while a sufficiently large in-plane momentum of the photon is maintained, such that  a significant portion of the Brillouin zone can be explored. Furthermore, one may also consider resonances induced by soft x-ray (e.g., the $2s \rightarrow 3d$ and the $3s \rightarrow 3d$ resonances in cuprates), which have larger wavefunction overlaps between the core orbitals and their nearest-neighbor valence orbitals.

\newcommand{\fnTwo}{The $3d$ orbitals in herbertsmithite is also rotated relative to the kagome plane.\cite{Shores:JACS:2005} While this would affect the precise values of the hopping magnitude, it should not affect the sign pattern as presented in the figure or the statement that the amplitudes are non-vanishing.}

For the rest of this section we shall assume that the signals from such inter-site transitions can indeed be detected and consider in detail whether the spin-chirality terms can indeed arise from such a case. Specifically, we shall focus on $s \rightarrow 3d$ inter-site transitions in Cu$^{2+}$ materials with the square and the kagome lattice geometries, having in mind the realistic materials of cuprates and herbertsmithite.  We shall also briefly comment on the cases of the triangular and the honeycomb lattices, in which the derivations of the spin-chirality terms are closely related to that of the square and the kagome lattices, respectively, and in which the former may be relevant to the new spin liquid candidate\cite{HDZhou:PRL:2011} Ba$_3$CuSb$_2$O$_9$. For brevity we shall drop the factors $g_i$ and $g_f$ that are common to all virtual processes.  

In such virtual processes with photon-induced inter-site hopping, it is easy to check that the intermediate state obtained after a photon-induced hop has an energy denominator of $\E_D = \delta\omega - (U - U_c) + i\Gamma$, where $\delta\omega = \omega_i - (\E_{3d} - \E_{\txt{core}})$ is again the detuning from the atomic transition. The resonant condition is thus given by $\omega_i \approx (\E_{3d} -\E_{\txt{core}}) + (U - U_c)$, under which $\E_D \approx i \Gamma$. \modified{For cuprates, $U \approx 8.8$~eV and $U_c \approx 7.0$~eV, while $\Gamma \approx 0.75$~eV for the \emph{$K$-edge}.\cite{Forte:PRB:2008} In comparison, in the effective one-band Hubbard model for cuprates, $t \approx 0.4$~eV.\cite{Hybertsen:PRB:1990}} Observe that the expression of $\E_D$ involves $U_c$, suggesting that the true resonant frequency of the inter-site dipole transition is offset from that of the intra-site quadruple transition by $U_c$. Indeed, the relative frequency shift between the inter-site dipole transition and the intra-site quadruple transition has been used to explain the ``three peaks'' feature of Ti pre-\textit{K}-edge absorption spectra in TiO$_2$.\cite{Uozumi:EPL:1992} The existence of such frequency shift may thus allow the signals from the intra-site quadruple transition to be further suppressed relative to inter-site dipole transition when the frequency of the incident photon is tuned.

\begin{figure}
\begin{center}
\subfigure[\label{fig:orientation_square}]{\includegraphics[scale=0.65]{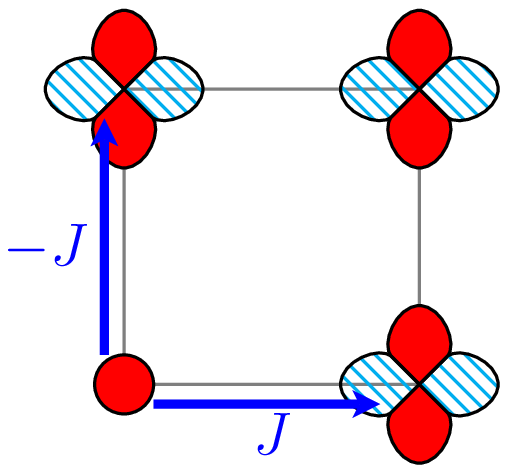}} \qquad
\subfigure[\label{fig:orientation_kagome}]{\includegraphics[scale=0.65]{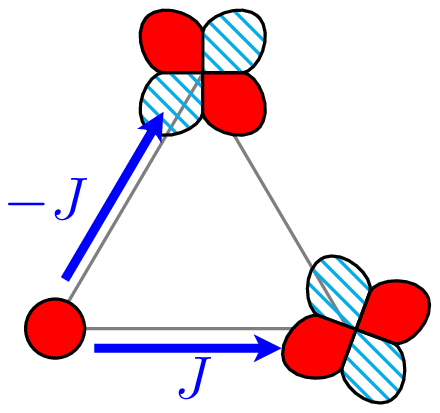}}
\caption{\label{fig:orientation} \colorfig Orientations of the $3d$ orbitals in (a) cuprates and (b) herbertsmithite, and their effects on the signs of the photon-induced hopping amplitudes in the $s \rightarrow 3d$ RIXS. Here the red (solid) and cyan (shaded) fillings indicate the relative signs of the angular part of the electron wavefunctions. Note from the figure that none of the photon-induced hopping magnitude is required to vanish by symmetry. }
\end{center}
\end{figure}

If we further assume that $\Gamma \ll U$ and that $U$ and $U_c$ are of the same order of magnitude, assumptions that appear to be valid in cuprates, then the virtual processes are dominated by those having intermediate states with exactly one low-energy valence doublon \emph{not located at the core-hole site}, and we can organize the $T$-matrix as an expansion in $t/\Gamma$ similar to that in the Raman case. 
Since the effect of the photon polarizations is now mostly reflected in the directions of the induced electron hops and since the spin-orbit coupling is negligible, we may take $H_C^{(1)} \sim \pm J (\vv{e}_{\alpha} \cdot \vvsym{\mu}) \sum_{\sigma} c^\dg_{i+\mu, \sigma} c_{i,s,\sigma} \equiv \pm J (\vv{e}_{\alpha} \cdot \vvsym{\mu}) (c^\dg_{i+\mu} c_{i,s})$ for the electron hop associated with the virtual absorption of photon and  $H_C^{(1)} \sim \pm J (\overline{\vv{e}}_{\alpha} \cdot \vvsym{\mu}) \sum_{\sigma} c^\dg_{i+\mu, s, \sigma} c_{i,\sigma} \equiv \pm J (\overline{\vv{e}}_{\alpha} \cdot \vvsym{\mu}) (c^\dg_{i+\mu, s} c_{i})$ for the electron hop associated with the virtual emission of photon, in which operators with the subscript $s$ correspond to the $s$ orbitals while operators without orbital labels correspond to the valence $3d$ orbitals. Note that $J$ is now a real scalar constant. The $\pm$ signs in the above equations are determined by the relative orientations of the $d$ orbitals and are illustrated\footnote{\fnTwo} in Fig.~\ref{fig:orientation}. 

\begin{figure}
\begin{center}
\subfigure[\label{fig:SSS_sq_1}]{\includegraphics[scale=0.8]{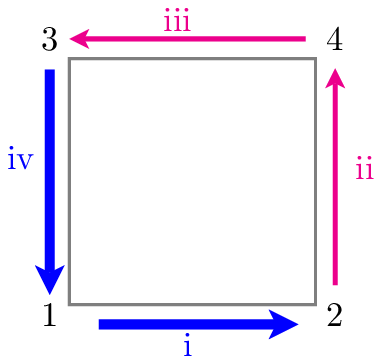}} \qquad
\subfigure[\label{fig:SSS_sq_2}]{\includegraphics[scale=0.8]{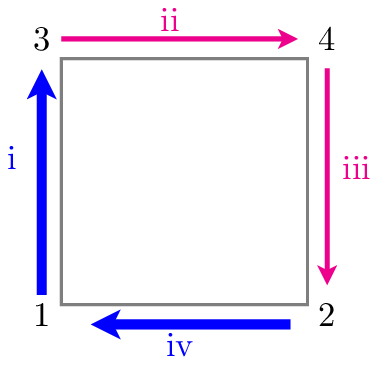}}
\caption{\label{fig:SSS_sq} \colorfig Two processes with inter-site photon-induced transitions that contribute to the spin-chirality terms in the $s \rightarrow 3d$ RIXS. }
\end{center}
\end{figure}

Now consider the particular case of the square lattice with only uniform nearest-neighbor hopping $t$. For such a lattice, virtual processes that involve valence electrons on at least three sites first appear at the order of two internal hops (i.e., at the ($t^2 J^2/\E_D^3$)-th order). Two such processes are depicted in Figs.~\ref{fig:SSS_sq_1} and \ref{fig:SSS_sq_2}. The corresponding contributions to the $T$-matrix are, respectively:
\begin{align}
T^{(a)}_{\txt{sq}} & = \frac{1}{\E_D^3} \cc[\ef{y} J]{1,s}{3} \cc[t]{3}{4} \cc[t]{4}{2} \cc[\ei{x} J]{2}{1,s} \notag \\
&= \frac{t^2 J^2}{\E_D^3} \efei{y}{x} \tr\{ \chi_3 \chi_4 \chi_2 \} \notag \\
& \doteq -\frac{2 i t^2 J^2}{\E_D^3} \efei{y}{x} \ \SSS{3}{4}{2} \notag \\
& = - \frac{2 i t^2 J^2}{\E_D^3} \efei{y}{x} \times \yytox \punct{,} \label{eq:SSS_sq_1} 
\end{align}
\begin{align}
T^{(b)}_{\txt{sq}} & =  \frac{1}{\E_D^3} \cc[-\ef{x} J]{1,s}{2} 
	\cc[t]{2}{4} \cc[t]{4}{3} \cc[-\ei{y} J]{3}{1,s} \notag \\
&= \frac{t^2 J^2}{\E_D^3} \efei{x}{y} \tr\{ \chi_2 \chi_4 \chi_3 \} \notag \\
& \doteq - \frac{2 i t^2 J^2}{\E_D^3} t^2 J^2 (\efei{x}{y}) \ \SSS{2}{4}{3} \notag \\
&  = \frac{2 i t^2 J^2}{\E_D^3} \efei{x}{y} \times \yytox \punct{,} \label{eq:SSS_sq_2}
\end{align}
where $\doteq$ denotes the part of the $T$-matrix that contains the spin-chirality terms, and that a graphic representation of the spin-chirality terms have been adopted on the fourth line of Eqs.~(\ref{eq:SSS_sq_1}) and (\ref{eq:SSS_sq_2}). 

For a fixed core-hole site, at this order there are three additional pairs of processes that contribute to the spin-chirality terms, which can be obtained from the processes depicted in Fig.~\ref{fig:SSS_sq} by successive $90^\circ$ rotations about the core-hole site. Summing over all these processes, to this order the contribution to the spin-chirality terms by a core hole at site $i$ is given by:
\begin{align}
T^{i}_{\txt{sq}} & \doteq \frac{2 i t^2 J^2}{\E_D^3}  (\efei{x}{y} - \efei{y}{x}) \notag \\
	& \quad \times \left( \ytoxx + \xxtoyy + \yytox + \xtoy \right) \punct{.} \label{eq:SSS_sq_site}
\end{align}
Summing over all possible core-hole sites, and now restoring the exponential factor $e^{i (\vv{k}_i - \vv{k}_f) \cdot \vv{r}_i}$, we have:\footnote{To be accurate, the position vector $\vv{r}$ in the exponential factor should be located at the bond center of the respective photon-induced hop. However, assuming that the momentum transferred is a fraction of the reciprocal lattice vector, it is permissible to neglect displacements that are only fractions of the lattice spacing.}
\begin{align}
T_{\txt{sq}} & \doteq \sum_{R} \frac{2 i t^2 J^2}{\E_D^3}  (\efei{x}{y} - \efei{y}{x}) e^{i (\vv{k}_i - \vv{k}_f) \cdot \vv{R}} \notag \\
& \quad \times \left( \ytoxx + \xxtoyy + \yytox + \xtoy \right)_\vv{R} \punct{,} \label{eq:SSS_sq_sum}
\end{align}
where the subscript $\vv{R}$ next to the parenthesis labels the site with which the spin-chirality terms are associated. From Eq.~(\ref{eq:SSS_sq_sum}) we see that for the square lattice there are indeed contributions to the $T$-matrix that couple to the spin-chirality terms at momentum $\vv{q} = \vv{k}_i - \vv{k}_f$ (i.e., the momentum transferred by the photon).

\begin{figure}
\begin{center}
\includegraphics[scale=0.8]{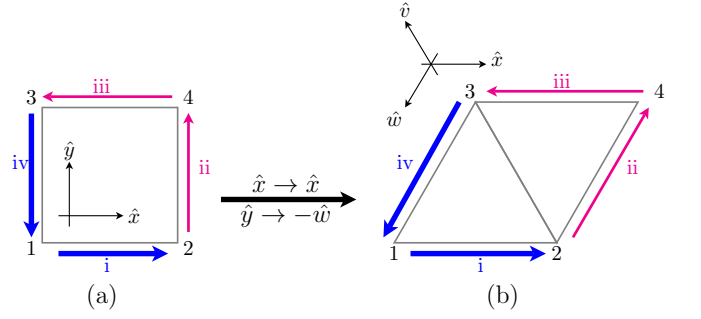}
\caption{\label{fig:SquareTriangular} \colorfig Mapping of two-internal-hop processes between the square and the triangular lattices. }
\end{center}
\end{figure}

The above analysis for the square lattice can be readily modified to the case of the triangular lattice, since processes with less than two internal hops can involve valence electrons at at most \emph{two} sites and hence do not give rise to any spin-chirality terms, while the two-internal-hop processes in the triangular lattice and the square lattice are topologically the same (see Fig.~\ref{fig:SquareTriangular} for illustration). For example, the contribution to the spin-chirality terms by the process depicted in Fig.~\ref{fig:SquareTriangular}(b) can be read off as:
\begin{equation}
T^{(b)}_{\txt{tri}} \doteq \frac{2 i t^2 J_i J_f}{\E_D^3}  \efei{w}{x} \dntri \punct{,} \label{eq:SSS_tri_ex}
\end{equation}
where the superscript $w$ in the photon polarization $\ef{w}$ corresponds to the unit vector $\uv{w}$ as depicted in the figure. Assuming that all photon-induced hops have the same amplitude, so that $J_i = J_f = J$ for all processes, we can sum up all contributions as in the square lattice case to obtain:
\begin{align}
T_{\txt{tri}} & \doteq \sum_{R} \frac{\sqrt{3} i t^2 J^2}{2 \E_D^3}  (\efei{x}{y} - \efei{y}{x}) e^{i (\vv{k}_i - \vv{k}_f) \cdot \vv{R}} \notag \\
& \quad \times \Bigg( 3 \uptri + 3 \dntri + \xtoww + \wwtov \notag \\
& \quad \quad + \vtoxx + \xxtow + \wtovv + \vvtox \Bigg)_\vv{R} \punct{.} \label{eq:SSS_tri_sum}
\end{align}
Hence, as in the square lattice case, the spin-chirality terms do appear in the triangular lattice at the ($t^2 J^2/\E_D^3$)-th order.

\begin{figure}
\begin{center}
\subfigure[\label{fig:SSS_kgm_fail_0}]{\includegraphics[scale=0.9]{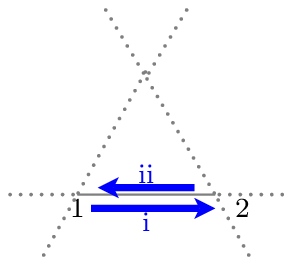}} \quad
\subfigure[\label{fig:SSS_kgm_fail_1}]{\includegraphics[scale=0.9]{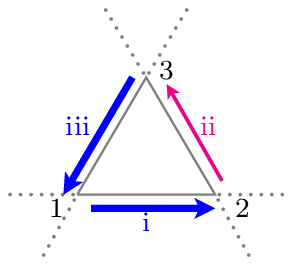}} \quad
\subfigure[\label{fig:SSS_kgm_fail_2}]{\includegraphics[scale=0.9]{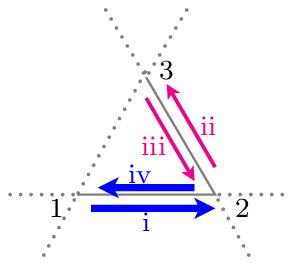}} \\
\subfigure[\label{fig:SSS_kgm_fail_3a}]{\includegraphics[scale=0.9]{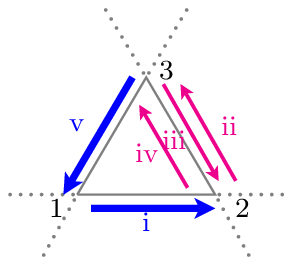}} \quad
\subfigure[\label{fig:SSS_kgm_fail_3b}]{\includegraphics[scale=0.9]{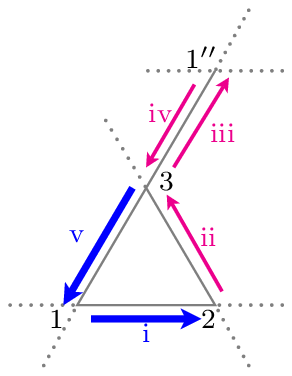}} \\
\subfigure[\label{fig:SSS_kgm_fail_3c}]{\includegraphics[scale=0.9]{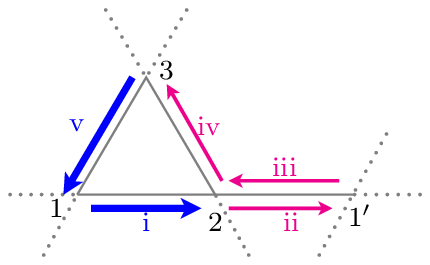}} \quad
\subfigure[\label{fig:SSS_kgm_fail_3d}]{\includegraphics[scale=0.9]{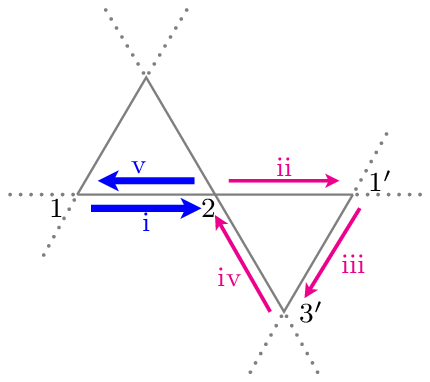}} 
\caption{\label{fig:SSS_kgm_fail} \colorfig Topologically distinct resonant virtual processes with inter-site photon-induced transitions in the $s \rightarrow 3d$ RIXS in the kagome lattice, up to three internal hops. }
\end{center}
\end{figure}

Next we consider the kagome lattice with only uniform nearest-neighbor hopping $t$. In Fig.~\ref{fig:SSS_kgm_fail} we list all the distinct \emph{topologies} (as opposed to geometries, such that, e.g., Fig.~\ref{fig:SSS_kgm_fail_2} is also representative of processes in which the third site is located at other locations) of resonant virtual processes up to three internal hops. Unfortunately, none of these processes generate any spin-chirality terms. To see this, first observe that the processes depicted in Figs.~\ref{fig:SSS_kgm_fail_0}--\ref{fig:SSS_kgm_fail_3a} involve valence electrons at only \emph{two or fewer} sites and thus cannot generate any spin-chirality terms (note that only \emph{core} electrons are involved at site 1 in all of the processes listed in Fig.~\ref{fig:SSS_kgm_fail}). Next, to rule out the processes depicted in Figs.~\ref{fig:SSS_kgm_fail_3b}--\ref{fig:SSS_kgm_fail_3d}, note that a spin state is annihilated by two successive creation operators or two successive annihilation operators on the same site, \emph{regardless of the spin characters of these two operators}. Consequently, if a site is transversed more than once in a virtual process, then each internal loop contributes to a separate trace in the Shastry--Shraiman derivation. For instance, corresponding to Fig.~\ref{fig:SSS_kgm_fail_3d} we have:
\begin{align}
T^{(g)}_{\txt{kagome}} & = \efei{x}{x} \frac{t^3 J^2}{\E_D^4} 
	\cc{1,s}{2} \cc{2}{3'} \cc{3'}{1'} \cc{1'}{2} \cc{2}{1,s} \notag \\
& = \efei{x}{x} \frac{t^3 J^2}{\E_D^4}  \cc{1,s}{2} \tr \{ \chi_{3'} \chi_{1'} \}  \cc{2}{1,s} \notag \\
& = \efei{x}{x} \frac{t^3 J^2}{\E_D^4}  \tr\{ \chi_{2} \} \tr \{ \chi_{3'} \chi_{1'} \} \notag \\
& = \efei{x}{x} \frac{t^3 J^2}{\E_D^4} \left( \frac{1}{2} + 2 \SdotS{3'}{1'} \right) \punct{.} \label{eq:SSS_kgm_fail_3g}
\end{align}
Similarly, $T^{(e)}_{\txt{kagome}} \sim ( 1/2 + 2 \SdotS{2}{3})$ and so does $T^{(f)}_{\txt{kagome}}$.

\begin{figure}
\begin{center}
\subfigure[\label{fig:SSS_kgm_1}]{\includegraphics[scale=0.9]{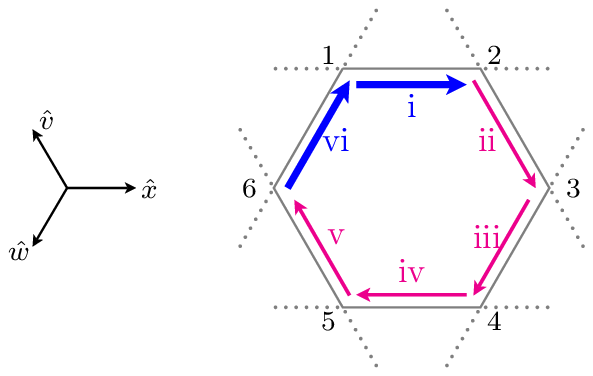}} \\
\subfigure[\label{fig:SSS_kgm_2a}]{\includegraphics[scale=0.9]{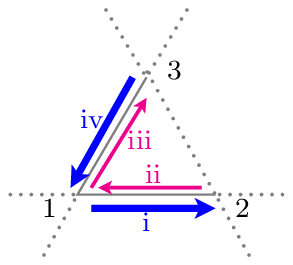}} \qquad
\subfigure[\label{fig:SSS_kgm_2b}]{\includegraphics[scale=0.9]{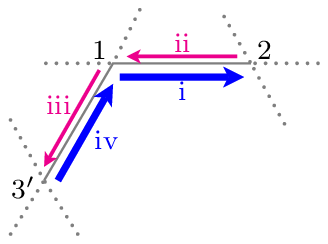}}
\caption{\label{fig:SSS_kgm} \colorfig Virtual processes with inter-site photon-induced transitions that contribute to the spin-chirality terms in the $s \rightarrow 3d$ RIXS in the kagome lattice. }
\end{center}
\end{figure}

For the kagome lattice, the spin-chirality terms first appear at the $(t^4 J^2/\E_D^5)$-th order, which arise from virtual processes in which the doublon hop through a hexagon. One such process is depicted in Fig.~\ref{fig:SSS_kgm_1}, whose contribution is given by:
\begin{align}
T^{(\txt{hex})}_{\txt{kagome}} & = 
	\efei{w}{x} \frac{t^4 J^2}{\E_D^5} \cc{1,s}{6} \cc{6}{5} \cc{5}{4} \cc{4}{3} \cc{3}{2} \cc{2}{1,s} \notag \\
	& = \efei{w}{x} \frac{t^4 J^2}{\E_D^5} \tr\{ \chi_6 \chi_5 \chi_4 \chi_3 \chi_2 \} \notag \\
	& \doteq \efei{w}{x} \frac{i t^4 J^2}{2 \E_D^5}
		\sum_{6 \geq a > b > c \geq 2} \vv{S}_{a} \cdot (\vv{S}_{b} \times \vv{S}_{c}) \punct{,} \label{eq:SSS_kgm_hex}
\end{align}
where the superscript $w$ in $\overline{e}^{w}_{f}$ corresponds to the unit vector $\hat{w}$ as depicted in the figure. One can check that processes with the same topology as the one depicted in Fig.~\ref{fig:SSS_kgm_1} sum to a nonzero contribution to the spin-chirality terms in the $(\efei{x}{y} - \efei{y}{x})$ channel at momentum $\vv{q}$ equal to that transferred by the photon, which for brevity we shall not write down explicitly. It can also be checked that any process at this order with a different topology does not contribute to any spin-chirality terms.

Compared with the corresponding terms in the square lattice, the spin-chirality terms in the kagome lattice are down by a factor of $(t/\E_D)^2$, which can be significant in the limit where $t \ll \Gamma$ even if the resonant condition is met. In such case one may want to consider also processes in which not all energy denominators are equal to $\E_D$. With this relaxed criterion, contributions to the spin-chirality terms can be found at the order of two internal hops, in which the doublon hops through the core-hole site (see Figs.~\ref{fig:SSS_kgm_2a} and \ref{fig:SSS_kgm_2b} for illustrations). For instance, the process depicted in Fig.~\ref{fig:SSS_kgm_2a} contributes:
\begin{align}
T^{(U_c)}_{\txt{kagome}} & 
	= -\efei{w}{x} \frac{t^2 J^2}{\E_D^2 (\E_D + U_c) } \cc{1,s}{3} \cc{3}{1} \cc{1}{2} \cc{2}{1,s} \notag \\
	& = -\efei{w}{x} \frac{t^2 J^2}{\E_D^2 (\E_D + U_c) } \tr\{ \chi_3 \chi_1 \chi_2 \} \notag \\
	& \doteq -\efei{w}{x} \frac{2 i t^2 J^2}{\E_D^2 (\E_D + U_c) } 
		\vv{S}_{3} \cdot (\vv{S}_{1} \times \vv{S}_{2}) \punct{.} \label{eq:SSS_kgm_Uc}
\end{align}
Again it can be checked that all processes with the same topology as the one depicted in Fig.~\ref{fig:SSS_kgm_2a} (which includes the one depicted in Fig.~\ref{fig:SSS_kgm_2b}) sum to a nonzero contribution to the spin-chirality terms in the $(\efei{x}{y} - \efei{y}{x})$ channel. It is worth noting that such ``back-tracking'' processes are also present in the square lattice and carry opposite signs from the ordinary ones depicted in Fig.~\ref{fig:SSS_sq}. Thus in the limit where $t \ll \Gamma$, the ratio of prefactors in the spin-chirality terms in the kagome lattice over that in the square lattice is given by $(\E_D + U_c)^{-1} / \left( \E_D^{-1} - (\E_D + U_c)^{-1} \right) = \E_D / U_c$. It is, however, worth noting that $t/\Gamma$ is not expected to be small in the case of cuprates.

Since the honeycomb lattice has the same hexagon loops as in the kagome lattice and has no shorter (in terms of the number of hops) loops, it can be readily checked that the spin-chirality terms again first appear in the honeycomb lattice at the $(t^4 J^2/\E_D^5)$-th order when $t \lesssim |\E_D|$ and at the $(t^2 J^2/ (\E_D + U_c)\E_D^2)$-th order when $t \ll |\E_D|$, with Figs.~\ref{fig:SSS_kgm_1} and \ref{fig:SSS_kgm_2b} the typical contributing processes in the respective cases.

\section{Discussions and Conclusions} \label{sect:discussions}

In this paper, we consider the question of whether RIXS can be used to detect many-body excitations that are coupled to the spin-chirality terms in a Mott insulator. We find that the spin-chirality terms are in general absent in the usual experimental setups, in which the spectroscopy is done near an absorption edge. The absence of the spin-chirality terms in these setups can be traced to the lack of linkage between the virtual electron hops and the photon polarizations.  However, we argue that RIXS still holds a prospect of observing the effects of the spin-chirality terms if one instead considers spectroscopy near a \emph{pre-edge}, in which case the intra-site dipole transitions are forbidden. 

Focusing on the Cu$^{2+}$ materials with the square and the kagome lattice geometries, we find that the spin-chirality terms are indeed presented  in both cases under our new proposal. However, in the kagome case such terms appear only at a higher order in our expansion. In addition, we also find that as far as the spin-chirality terms are concerned, the scenario for the triangular lattice is analogous to that of the square lattice, while the scenario for the honeycomb lattice is analogous to that of the kagome lattice. It is worth noting that the situation we encounter in RIXS is essentially the reverse of what happens in the Raman case, in which the spin-chirality terms occur at the ($t^4/U^3$)-th order in the kagome and the honeycomb lattices but not in the square or the triangular lattices.

In comparison to the similar scheme to detect the spin-chirality terms in Raman spectroscopy,\cite{Shastry:PRL:1990,Shastry:IJMPB:1991} which had already been realized,\cite{Sulewski:PRL:1991} the present scheme in RIXS suffers from the reduced wavefunction overlaps in the inter-site dipole transitions. However, it has the advantage that excitations with finite momentum can be probed. To put this into perspective, let us return to the motivation we presented in the introduction, namely the emergent gauge boson in the $U(1)$ Dirac spin liquid. In the $U(1)$ Dirac spin liquid, the spin-chirality terms in the $T$-matrix correspond to flux-flux correlators, viz.:
\begin{align}
\sum_f W_{fi} & = \sum_f 2\pi |\bra{f} T \ket{i}|^2 \delta(\E_f - \E_i) \\
&	\sim \bra{i} b(\vv{\Delta k}, \Delta\omega) b(\vv{0}, 0) \ket{i} + \cdots \notag \\
& \propto \frac{q^2 \Theta(\Delta\omega - v_F \Delta k)}%
	{(\Delta\omega^2 - v_F^2 \Delta k^2)^{1/2}} + \cdots \punct{,} \label{eq:spectral}
\end{align}
where $\Theta$ denotes the step function, $v_F$ is the Fermi velocity at the Dirac cone of the $U(1)$ Dirac spin liquid, $b$ is the ``magnetic field'' associated with the emergent gauge boson, and $\Delta\omega = \omega_i - \omega_f$ ($\vv{\Delta k} = \vv{k}_i - \vv{k}_f)$ is the energy (momentum) transferred from the photon. If we assume that $\E_D$ in RIXS and $(\omega_i - U)$ in Raman spectroscopy are of the same order, the intensity of the signal from the gauge boson in RIXS will be modified from that in Raman spectroscopy by a factor roughly equal to $J^2 \E_D/t^2 (\E_D + U_c)$ or $J^2/\E_D^2$, depending on which limit one considers in RIXS. However, such comparison is not particularly meaningful since we have not considered how the \emph{background} signals compare in the two cases. However, the advantage offered by RIXS is not so much in the intensity of the signal but rather in its lineshape. In the Raman case where $\vv{\Delta k} \approx 0$, the signature of the emergent gauge boson can manifest only as a power-law behavior near zero energy transfer, which can easily be masked by the elastic or quasielastic peak. In contrast, in RIXS the signal from the gauge boson has a sharp threshold at $\Delta\omega = v_F \Delta k$, which varies as $\vv{\Delta k}$ varies. Thus, assuming modest intensities of the signals, it would be much easier to discern the emergent gauge boson in the case of RIXS.

Of course, one should not underestimate the experimental challenges in realizing the proposal laid out in this paper. However, enormous progress in RIXS has been made in recent decades,\cite{Kotani:RMP:2001} with two-magnon excitations being observed\cite{Hill:PRL:2008, Braicovich:PRL:2009} and three-magnon excitations being proposed. \cite{Ament:arXiv:1002.3773} It is our hope that our proposal will further stimulate new theoretical and experimental advances in the field.

\begin{acknowledgments}
We thank George Sawatzky, Akio Kotani, and Peter Abbamonte for helpful information. This research was supported in part by the DOE under Grant No.\@ DE-FG02-03ER46076 (W.H.K and P.A.L.) and in part by the NSF under Grant No.\@ NSF PHY05-51164 (W.H.K).
\end{acknowledgments}



\end{document}